\documentclass[useAMS,usenatbib]{mn2e}

\def\Term#1 #2 #3/{\mbox{$^{#1}\!#2_{#3}$}}
\def\Termo#1 #2 #3/{\mbox{$^{#1}\!#2^o_{#3}$}}
\def\sterm #1 #2 #3/{\mbox{$_{#3}\!^{#1}\!#2$}}
\def\qP{$^4\!P$}
\def\qD{$^4\!D$}
\def\qG{$^4\!G$}
\def\qF{$^4\!F$}

\def\dI{$^2\!I$}
\def\dH{$^2\!H$}
\def\dF{$^2\!F$}
\def\dP{$^2\!P$}

\begin{document}

\title[MCDHF energy levels and transition probabilities for $3d^5$ in Fe~\sc{iv}]
{Multiconfiguration Dirac-Hartree-Fock energy levels and transition 
probabilities for $3d^5$ in Fe~\sc{iv}}

\author[C. Froese Fischer, R. H. Rubin, and M. Rodr\'iguez]{C. Froese Fischer$^1$\thanks{E-mail:
charlotte.fischer@nist.gov}, R. H. Rubin$^{2,3,4}$, and M. Rodr\'iguez$^5$ \\
$^1$National Institute of Standards and Technology,  Gaithersburg, MD
20899-8422, USA \\
$^2$NASA Ames Research Center, Moffett Field, CA 94035-1000, USA\\
$^3$Orion Enterprises, M.S. 245-6, Moffett Field, CA 94035-1000,USA\\
$^4$Kavli Institute for Astronomy and Astrophysics, Peking University,
Beijing 100871, P. R. China\\
$^5$Instituto Nacional de Astrof\'\i sica, \'Optica y Electr\'onica,
Apdo Postal 51 y 216, 72000 Puebla, Mexico}

\pagerange{\pageref{firstpage}--\pageref{lastpage}} \pubyear{2008}

\maketitle

\label{firstpage}

\begin{abstract}
Multiconfiguration Dirac-Hartree-Fock electric quadrupole (E2) and magnetic
dipole (M1)  transition probabilities are reported for
transitions between levels of $3d^5$ in [{Fe~\sc{iv}}]. 
The accuracy of the {\sl ab initio} energy levels
and the agreement in the length and velocity forms of the line strength for the
E2 transitions are used as indicators of accuracy.
The present  E2 and M1 transition probabilities are compared with earlier
Breit-Pauli results and other theories.
An extensive set of transition probabilites with indicators of accuracy are
reported in Appendices A and B. Recommended values of $A$(E2) + $A$(M1) are listed in 
Appendix C. 

\end{abstract}


 
\section{Introduction}
Iron is one of the main contributors to the mass of refractory dust
grains~\citep*{sofia}.
The determination of its gaseous abundance in ionized
nebulae of different characteristics can be used to infer the efficiency of dust
destruction and formation processes in widely different environments, ranging
from H~{\sc ii} regions and their associated molecular clouds to planetary nebulae
(PNe), the descendants of asymptotic giant branch stars, 
which are the main source of dust grains in
the solar neighbourhood~\citep{whittet2003}.

In these photoionized nebulae, Fe$^{3+}$ is an important ionization state,
being the dominant ion in most H~{\sc ii} regions and many PNe. Several
{[Fe~\sc{iv}]} lines are expected to appear in the ultraviolet (UV), 
optical, and infrared spectra of these objects
but, unfortunately, the more easily accessible optical lines are very weak and
difficult to measure. Hence, the first {[Fe~\sc{iv}]} line observed in an
H~{\sc ii} region
was the UV {[Fe~\sc{iv}]} 2836.56~\AA\ line, observed by ~\citet{rubin1997} in
the Orion Nebula with the Hubble Space Telescope. Rubin {et al.} found that the
Fe$^{3+}$ abundance implied by this line was much lower than expected.
Subsequent observations of this and some weak optical {[Fe~\sc{iv}]} lines in a
handful of H~{\sc ii} regions and PNe confirmed the discrepancy (see
\citet{RR2005} and references therein). Since the discrepancy translates into a
large uncertainty in the iron abundances calculated for most nebulae, it
stresses the need for reliable atomic data for the Fe ions, especially the
transition probabilities and electron impact excitation collision strengths
that are needed to solve the equations of statistical equilibrium for the lower
energy levels of these ions~\citep{OF2006}.

Because of the rough scaling of the line intensity with the product
of the Fe$^{3+}$ density and the electron density, optical {[Fe~\sc{iv}]} lines
are more easily measured in objects with
relatively large densities ($N_e\approx10^6$ cm$^{-3}$), like some
PNe~\citep*{Rodriguez,zhang2002,Zhang2005}. However, these
spectra are so cluttered with lines that line
identification and deblending can be a problem. This is another instance where
reliable atomic data are needed in order to provide good estimates of 
the relative line intensities.

The first values of radiative transition rates for {[Fe~\sc{iv}]} were calculated as
early as 1958 by Garstang,
who used them to confirm the identification by
\citet{thackeray} of several {[Fe~\sc{iv}]} lines in the spectrum of the symbiotic nova
RR~Telescopii. Forty years later, \citet{rubin}, motivated by
the observation and analysis of {[Fe~\sc{iv}]} 2836.56~\AA\ in the Orion Nebula by
\citet{rubin1997}, provided improved values for the transition probabilities
between the 12 lowest levels of Fe$^{3+}$, the same levels for which effective
collision strengths had just become available 
\citep{berrington1995,berrington1996}.
However, most of the observed {[Fe~\sc{iv}]} lines arise from higher energy levels, and
new calculations that included these levels were not long in coming: collision
strengths from \citet{zhang1997}  and transition probabilities 
from \citet{bp}.

The goal of the first \citet{rubin} publication
was to predict radiative transition probabilities from $3d^5$ \Term 4 G {}/,
\Term 4 P {}/, and \Term 4 D {}/ levels to the \Term 6 S {5/2}/ ground state.
In the process, transition probabilities were also computed for transitions between the levels of
the quartet terms.
Four theoretical methods were compared and best estimates identified.
The multiconfiguration Dirac-Hartree-Fock (MCDHF) with the Breit correction
was selected in some cases for
M1 transitions but there was a strong indication of an error in the code
for some E2 transitions.
In the second paper~\citep{bp} the multiconfiguration Hartree-Fock
(MCHF) work with  Breit-Pauli corrections was extended
to include all levels of the
$3d^5$ configuration. In these later calculations,  term
energy corrections were used to adjust the position of a term to be in close
agreement with the values cited in the
Atomic Spectra Database (ASD)~\citep{asd}, thereby improving the wave function
$LS$ term composition.
Because there are as many as three terms with the same $LS$ value, as for the
\Term 2 D {}/  (\sterm 2 D 5/, \sterm 2 D 3/,
and \sterm 2 D 1/ where the
preceding subscript is the seniority), the adjustments were done in groups
in order of energy. Thus the final energies are semi-empirical and only the
fine-structure splitting provides an indication of accuracy.  In a more
recent paper, \citet{nahar} reported radiative transition rates for
a large number of lines that include electric dipole (E1), quadrupole (E2),
octupole (E3) and magnetic dipole (M1) transitions in {Fe~\sc{iv}}. The calculations
included only 
the one-body relativistic corrections  of the Breit-Pauli Hamiltonian.
Transition calculations were based on the {\sl ab initio}
line strength and ASD transition energies.

Assessing the accuracy of theoretical results in the absence of reliable 
experimental data is as difficult (if  not more difficult) than the 
transition calculations themselves.  One way of establishing accuracy is
by validating the results through calculations based on different
theories.
In this paper we report results for fully {\sl ab initio} multiconfiguration
Dirac-Hartree-Fock results where relativistic effects are included
in the basic theory and not added as a low-order correction as in the Breit-Pauli
calculations by~\citet{bp} and \citet{nahar}.  Computed
energy levels and their fine-structure splitting can be used as an indicator
of the accuracy of the wave function along with  the agreement in
the length and velocity forms of the line strength for E2 transitions.
All calculations were performed using the most recent
revised and corrected parallel GRASP2K code \citep{grasp2K}.  

\section{Computational Procedure }
In the MCDHF approach \citep{GRANT} the wave function $\Psi$
for a state labelled $\gamma J$,
 where $\gamma$ represents the configuration
and any other quantum numbers
 required to specify the state, is approximated by an expansion
over $jj$-coupled configuration state
functions (CSFs) 
\begin{equation}
\Psi(\gamma J)=\sum_{j}c_{j}\Phi(\gamma_{j}J).
\end{equation}
The configuration state functions $\Phi(\gamma_j J)$
are anti-symmetrized linear
combinations of products of  relativistic orbitals
\begin{equation}
\phi({\bf r}) = \frac{1}{r}
 \left( \begin{array}{c}
P_{n \kappa}(r) {\chi}_{\kappa m}(\hat{r}) \\
iQ_{n \kappa}(r) {\chi}_{-\kappa m}(\hat{r})  \end{array} \right).
\end{equation}
Here  $\kappa$ is the
relativistic angular quantum number,
 $P_{n \kappa}(r)$ and $Q_{n \kappa}(r)$
 are the large and small component radial
wave functions and ${\chi}_{\kappa m}(\hat{r})$
is the spinor spherical harmonic
in the $lsj$ coupling scheme
\begin{equation}
{\chi}_{\kappa m} (\hat{r})=
 \sum_{m_l,m_s}\langle l \frac{1}{2}m_l m_s \mid 
jm \rangle Y_{lm_l}(\theta,\varphi) \xi_{m_s}(\sigma).
\end{equation}
The radial functions $P_{n \kappa}(r)$ and $Q_{n \kappa}(r)$
are numerically represented 
on a logarithmic grid and are required to be
orthonormal within each $\kappa$ symmetry,
\begin{equation}
\int_{0}^{\infty}[P_{n^{\prime}\kappa}(r)P_{n \kappa}(r)
+ Q_{n^{\prime} \kappa}(r)Q_{n \kappa}(r)]dr = \delta_{n^{\prime}n}.
\end{equation}
In the multiconfiguration self-consistent
 field (MC-SCF) procedure both the
radial functions and the expansion coefficients for
the configuration state functions are optimized to self-consistency.

Once a set of radial orbitals has been obtained, relativistic
configuration interaction (RCI) calculations
can be performed to include the 
Breit interaction and quantum electrodynamic (QED) effects. 
At this stage only the expansion coefficients of the CSFs
are determined. This is done by diagonalizing the Hamiltonian matrix.

 In the relativistic configuration interaction
 calculations the transverse photon interaction
\vspace{1in}
\vspace{-1in}
\begin{eqnarray}
{\cal H}_{trans}& = & - \sum_{i<j}^N \left[
\frac{{\mbox{\boldmath $\alpha$}}_i
\cdot{\mbox{\boldmath
$\alpha$}}_j~ \mbox{cos} (\omega_{ij}r_{ij})}{r_{ij}}\right.  \nonumber \\
   & & \quad  + \left. 
( {\mbox{\boldmath $\alpha$}}_i \cdot \nabla_i)
( {\mbox{\boldmath $\alpha$}}_j \cdot \nabla_j)
\frac{\mbox{cos}(\omega_{ij}r_{ij})-1}{\omega_{ij}^2 r_{ij}} \right]
\end{eqnarray}
may be included in the Hamiltonian.
The photon frequency $\omega_{ij}$ used by the RCI program
in calculating the matrix elements
of the transverse photon interaction
is taken to be the difference in the diagonal Lagrange multipliers
${\epsilon_i}$ and ${\epsilon_j}$ associated with the orbitals.
In general, diagonal Lagrange multipliers
 are approximate electron removal
energies only when orbitals are spectroscopic and singly
occupied.  Thus it is
not known how well
 the code can determine the full transverse photon
interaction when correlation
orbitals are present and orbitals are multiply occupied as in the
present case.
 What can be obtained instead is the low frequency limit
$\omega_{ij} \rightarrow 0$
 usually referred to as the Breit interaction.

The transition parameters, such as rate and weighted
oscillator strength, for a multipole transition of rank $K$ from
$\gamma J$ to $\gamma^{\prime}J^{\prime}$ are all related to the
reduced transition matrix element
\begin{equation}
\langle \Psi(\gamma
J)\|{\bf O}^{(K)} \|\Psi(\gamma^{\prime}J^{\prime})
\rangle,
\end{equation}
where ${\bf O}^{(K)}$ is
 the relevant transition operator \citep{GRANTTP}.  In the present
 study these are the E2 and M1 transitions between levels of $3d^5$ 
but with transitions
 to the $3d^5\;^6S_{5/2}$ ground state being of greatest interest.

\section{MCDHF Calculation and Results}

\begin{table*}
 \centering
 \begin{minipage}{140mm}
\caption{Computed energy levels (in cm$^{-1}$), splitting
(separation between levels of a term), difference from
observed~\citep{asd} (computed - observed), and composition of the
$3d^5$ levels. The $J$-values within a term that are not in the observed
order are preceeded by an asterisk.}
\label{spectrum}
\begin{tabular} {@{} r r r r r r r l @{}}
  \hline
\multicolumn{1}{c}{level}
& \multicolumn {1}{l}{LS}
&\multicolumn {1}{c}{J}
&\multicolumn {1}{c}{Energy}
&\multicolumn {1}{c}{Splitting}
&\multicolumn {1}{c}{\quad Diff.}
&\multicolumn {1}{c}{\ }
&\multicolumn {1}{c}{Composition (\%)}\\
 \hline
1 & ${\!^6\!S}$ & 5/2    &      0.0  &     0.0 &     0.0  && 97\\
2 &${\!^4\!G}$ & 11/2    &  33491.9  & &  1246.4  && 96\\
3 &            & 9/2     &  33538.1  &    46.2 &  1245.3  && 97  \\
4 &            &  5/2    &  33547.3  &     9.2 &  1246.1  && 97\\
5 &            &  7/2    &  33551.6  &     4.3 &   1245.9  && 97\\
6 &${\!^4\!P}$ &  5/2    &  35721.3  &  &   467.5  && 93 + 3~\sterm 4 D {}/\\
7 &            &   3/2   &  35791.4  &    70.2 &   458.1  && 94 + 2~\sterm 4 D {}/\\
8 &            &   1/2   &  35850.1  &    58.7 &   443.5  && 96 \\
9 & ${\!^4\!D}$&  7/2    &  39837.1  &   &  1057.7  && 96\\
10 &           &   1/2   &  39945.5  &   108.4 &  1048.8  && 96 \\
11 &           &   5/2   &  39963.9  &    18.4 &  1028.8  && 94 + 3~\sterm 4 P {}/\\
12 &           &   3/2   &  39973.2  &     9.2 &  1035.0  && 94 + 2~\sterm 4 P {}/\\
13& ${\!^2\!I}$&  11/2   &  49190.0  &  &  2099.5  && 96\\
14 &           &   13/2  &  49202.9  &    12.8 &  2112.4  && 96 \\
15 & $_5{\!^2\!D}$ & 5/2 &  50558.7  &  &  1017.2  && 55 + 22~\sterm 2 F 3/ + 18~\sterm 2 D 1/\\
16 &               & 3/2 &  51043.7  &   484.9 &   992.3  && 70 + 23~\sterm 2 D 1/ + 3~\sterm 4 F {}/\\
17 & $_3{\!^2\!F}$ & 7/2 &  52442.9  &  &  1049.7 && 94 + 1~\sterm 4 F {}/ +  1~\sterm 2 F 5/\\
18 &               & 5/2 &  53192.8  &   749.9 &  1026.1  && 70 + 7~\sterm 4 F {}/  + 14~\sterm 2 D 5/ + 4~\sterm 2 D 1/\\
19 & ${\!^4\!F}$ &   9/2 &  53833.1  &   &  1212.4  && 95  + 1~\sterm 2 G {}/\\
20 &             &   7/2 &  53896.4  &    63.3 &  1201.0  && 95 + 1~\sterm 2 F 3/\\
21 &             &  *3/2 &  54022.3  &  125.9  &  1185.2  && 93 +  3~\sterm 2 D 5/ + 1~\sterm 2 D 1/\\
22 &             &  *5/2 &  54007.1  &   -15.2 &  1169.1  && 88 + 4~\sterm 2 F 3/ + 3~\sterm 2 D 5/\\
23 &  ${\!^2\!H}$&   9/2 &  57734.1  &  &  1675.8  && 84 +  12~\sterm 2 G {}/\\
24 &             &  11/2 &  58007.6  &   273.5 &  1638.8  && 96\\
25 & $_5{\!^2\!G}$ & 7/2 &  59234.6  &  &  1826.6  && 96 \\
26 &               & 9/2 &  59504.8  &   270.3 &  1783.6  && 83 + 12~\sterm 2 H 3/ + 1~\sterm 4 F {}/\\
27 & $_5{\!^2\!F}$ & 5/2 &  62901.4  &  &  1744.8  && 95\\
28 &               & 7/2 &  62999.2  &    97.8 &  1744.8  && 96 + 1~\sterm 2 F 3/\\
29 & ${\!^2\!S}$   & 1/2 &  68332.5  &  &  1612.3  && 96\\
30 & $_3{\!^2\!D}$ & 3/2 &  75648.8  &  &  1552.2  && 96\\
31 &               & 5/2 &  75686.5  &    37.7 &  1573.4  && 96\\
32 &$_3{\!^2\!G}$  &  9/2&  85049.1  &  & 2154.3  && 96 \\
33 &               &  7/2&  85050.3  &     1.1 &  2153.0  && 96\\
34 & ${\!^2\!P}$ & 3/2   & 102444.0  & & 2326.0  && 96 \\
35 &           &   1/2   & 102447.9  &     3.9 &  2321.9  && 96 \\
36 &$_1{\!^2\!D}$ &5/2   & 110374.1  &  &  2122.0  && 72 + 23~\sterm 2 D 5/\\
37 &              & 3/2  & 110392.0  &    17.9 &  2133.7  && 72 + 23~\sterm 2 D 5/\\
\hline
\end{tabular}
\end{minipage}
\end{table*}

A series of calculations were performed that will be identified by the maximum
principal quantum number $n=3,4,5, 6$ of orbitals of the wave function. 
In all cases, the $1s^22s^22p^6$ 
core was treated as an inactive core.
In our first $n=3$ calculation the wave function expansion
included all single- and double- excitations
to the $3d$ subshell or 
CSFs obtained from $3s\rightarrow 3d$ and $3p^2 \rightarrow
3d^2$ excitation.  All orbitals were varied in an extended optimal level (OL)
calculation for all the 37 lowest levels: four for $J=1/2$, seven for $J=3/2$,
ten for $J=5/2$,
seven for $J=7/2$, five for $J=9/2$, three for $J=11/2$, and one for $J=13/2$. 
 This calculation accounts for the
near degeneracy effects between CSFs in the $n=3$ shell.

The $n=4$ calculation included an additional layer of orbitals, namely
$\{4s, 4p, 4d, 4f\}$ orbitals with a wave function expansion that, 
in addition to the $n=3$ CSFs,
included single- and double- (SD)
excitations from $3p^63d^5$, but with at most one excitation
from $3p^6$ as in a core valence calculation, with $3p^6$ being
part of the core and $3d^5$  the valence configuration.  
The latter included 37 486 CSFs. Only the $4l$ orbitals were varied.  
Since these orbitals are not occupied in the single configuration 
wave function, these orbitals are {\it virtual} orbitals also referred
to as correlation orbitals.  The $n=5$ calculation extended the
correlation orbitals to include also $\{5s, 5p, 5d, 5f, 5g\}$ orbitals. The 
SD excitation process (restricted to at most one excitation from $3p^6$) 
resulted in a wave function expansion of 213 037 CSFs for our range of $J$.
The final $n=6$ calculation introduced \{$6s, 6p, 6d, 6f, 6g$\} orbitals but the
only CSFs added to the expansion were those from SD excitations from
$3d^5$. Because of the importance of the $J=5/2$ levels in this work, this
last set of orbitals were optimized for the ten $J=5/2$ levels.  Once
optimized radial functions were determined, relativistic configuration
interaction (RCI) calculations were performed to include the Breit
correction. The sizes of the matrices were $18\,350,\;33\,356,\; 42\,642,\; 45\,325,\;
42\,221,\; 35\,117,$ and $26\,359$ respectively for J = 1/2 to $J=13/2$.  

In this description, we have used a non-relativistic terminology, but the
JJGEN program \citep*{jjgen} used to generate the CSFs translates the terminology
to the relativistic framework where a $4p$ orbital, for example, is either
a $4p_{1/2}$ or $4p_{3/2}$ orbital and the coupling of each CSF is described in terms
of $jj$-coupling.

\begin{table}
\caption{Comparison of $J=5/2$ energy level separation (in cm$^{-1}$) with observed.  }
\label{separation}
\centering
\begin{tabular}{@{} r r r r r c @{}}
\hline
Term & Level & \ & \multicolumn{2}{c}{Separation} & ratio \\
\cline{4-5}
     & (Obs) &\ &   Obs &  Calc& (Calc/obs)\\
\hline
\Term 6 S {}/ & 0 \\
\Term 4 G {}/ & 32301 && 32301 & 33547 & 1.04 \\
\Term 4 P {}/ & 35254 && 2953 &  2174 &  0.74\\
\Term 4 D {}/ & 38935 && 3681 &  4243 & 1.15\\
\sterm 2 D 5/ & 49542 && 10606 &  10595& 1.00\\
\sterm 2 F 3/ & 52167 && 2625 & 2634 & 1.00\\
\Term 4 F {}/ & 52838 && 671 & 814 & 1.21 \\
\sterm 2 F 5/ & 61157 && 8319 & 8894 & 1.07\\
\sterm 2 D 3/ & 74133 && 12977 & 12785 & 0.99\\
\sterm 2 D 1/ & 108242 && 34109 & 34687 & 1.02\\
\hline
\end{tabular}
\end{table}

Since the present calculations are entirely {\sl ab initio}, 
our measure of accuracy 
will be the energy level structure of terms and their fine-structure. 
 The $3d^5\;^6S$ ground state, where
all spin-quantum numbers are the same, has some correlation in the motion
of the electrons included already at the single configuration
Dirac-Hartree-Fock (DHF) level through the antisymmetry 
requirement of the CSF.  
In fact the correlation correction to the total energy
 for the $^6S$ term is much smaller than for other
terms.  Thus we cannot expect the higher energy levels relative to $^6S$ to
be in good agreement with observed levels~\citep{asd}, but they should be in better 
relative
agreement with each other.  In particular the difference between
the computed and observed level structure should be essentially constant
within a term.

The final ($n=6$) energy levels relative to the ground state for these 
calculations are 
reported in Table \ref{spectrum}. The energy levels of the lowest term ($^4G$) 
differ from 
observed by 3.7~\% but those of the highest term (\sterm 2 D 1/) are accurate 
to 2.0~\% and compare favourably with errors of more than 10~\% for lower and
5~\% for the higher levels reported by ~\citet{nahar}.
The MCDHF energies include 
the Breit correction which is important in changing the order of the levels
for the $^4G$ term 
from [5/2, 7/2, 9/2, 11/2] to the observed order of [11/2, 9/2, 5/2, 7/2].
In addition the energies
relative to the lowest (\Term 4 G {11/2}/) of 
[46.2, 56.0, 60.0] cm$^{-1}$ are in excellent
agreement with observed values of [47.3, 55.7, 60.2]~\citep{asd}, respectively. 
The only fine-structure levels not in their observed order are the
$J=5/2, 3/2$ levels of \Term 4 F {}/: our calculated values are separated
by 15.2 cm$^{-1}$ whereas the observed $J=3/2$ should be lower than $J=5/2$
by 0.9 cm$^{-1}$.  However this does not directly contribute to the 
error in the wave function since only CSFs of the same parity and $J$
have non-zero interaction matrix elements. Comparing the wave function
composition to the earlier MCHF Breit-Pauli term composition, there is a 
general reduction in the admixture of different $LS$ terms.  For example,
whereas the present $_3$\Term 2 F {5/2}/ has a dominant contribution of 70~\%
it was only 60~\% in the earlier work.

\begin{table*}
 \centering
 \begin{minipage}{140mm}
\caption{ Observed vacuum wavelength $\lambda$ (\AA) and calculated
 line strength $S$, weighted absorption oscillator
strength $gf$, and radiative transition rate $A_{ki}$ (s$^{-1}$)
for E2 and M1 transitions to
the ground state. Transition rates normalized to observed
wavelengths ($A_{ki}$(N)) and unnormalized {\sl ab initio} values
($A_{ki}$(U)) are
reported. Also included are the accuracy indicators: $\delta E = 
\Delta E(calc)/\Delta E(obs) - 1 $ and
$\delta T = |S(\mbox{length})-S(\mbox{velocity})|/ \max(S(\mbox{length}), S(\mbox{velocity}))$. 
}
\label{excitation}
\begin{tabular}{@{}r r r l l l l l l l @{}}
  \hline
Term &  $J$& $\lambda$(vac) & & $S$ & $gf$  & $A_{ki}$(N) &
$A_{ki}$(U) &  $\delta E$ & $\delta T$ \\
 \hline
${\!^4\!G}$&  9/2&    3096.67& E2 &      2.361(-08) & 1.335(-16) & 9.285(-09) & 1.122(-08) & 0.037 & 0.038\\
           &  5/2&    3095.86& E2 &      2.033(-08) & 1.150(-16) & 1.334(-08) & 1.612(-08) & 0.037 & 0.066\\
           &     &           & M1 &      1.400(-07) & 1.828(-13) & 2.120(-05) & 2.375(-05) & \\
           &  7/2&    3095.43& E2 &      2.417(-08) & 1.368(-16) & 1.190(-08) & 1.438(-08) & 0.037 & 0.109\\
           &     &           & M1 &      1.883(-10) & 2.460(-16) & 2.141(-08) & 2.398(-08) & \\
${\!^4\!P}$&  5/2&    2836.57& E2 &      3.862(-05) & 2.841(-13) & 3.925(-05) & 4.192(-05) & 0.013 & 0.016\\
           &     &           & M1 &      6.642(-03) & 9.468(-09) & 1.308      & 1.361 \\
           &  3/2&    2830.19& E2 &      1.484(-05) & 1.099(-13) & 2.288(-05) & 2.441(-05) & 0.013 & 0.031\\
           &     &           & M1 &      2.855(-03) & 4.080(-09) & 8.494(-01) & 8.829(-01) & \\
${\!^4\!D}$&  7/2&    2578.69& E2 &      8.962(-04) & 8.776(-12) & 1.100(-03) & 1.259(-03) & 0.027 & 0.077\\
           &     &           & M1 &      1.565(-06) & 2.455(-12) & 3.078(-04) & 3.337(-04) & \\
           &  1/2&    2570.91& E2 &      3.499(-05) & 3.458(-13) & 1.828(-04) & 2.088(-04) & 0.018 & 0.033\\
           &  5/2&    2568.38& E2 &      6.436(-04) & 6.378(-12) & 1.075(-03) & 1.225(-03) & 0.026 & 0.090\\
           &     &           & M1 &      1.240(-04) & 1.953(-10) & 3.291(-02) & 3.559(-02) & \\
           &  3/2&    2568.17& E2 &      2.518(-04) & 2.496(-12) & 6.312(-04) & 7.196(-04) & 0.026 & 0.089\\
           &     &           & M1 &      3.969(-05) & 6.250(-11) & 1.580(-02) & 1.710(-02) & \\
$\,_5\,{\!^2\!D}$&  5/2 & 2018.51& E2 & 1.979(-08) & 4.040(-16) & 1.102(-07) & 1.220(-07) & 0.020 & 0.175\\
           &     &           & M1 &      1.699(-06) & 3.404(-12) & 9.287(-04) & 9.871(-04) & \\
           &  3/2&    1997.95& E2 &      9.416(-10) & 1.982(-17) & 8.281(-09) & 9.135(-09) & 0.019 & 0.402\\
           &     &           & M1 &      1.278(-07) & 2.587(-13) & 1.081(-04) & 1.146(-04) & \\
$\,_3\,{\!^2\!F}$&  7/2& 1945.74& E2 &  2.233(-07) & 5.090(-15) & 1.121(-06) & 1.240(-06) & 0.020 & 0.080\\
           &     &           & M1 &      5.087(-11) & 1.057(-16) & 2.328(-08) & 2.474(-08) & \\
${\!^4\!F}$&  9/2&    1900.39& E2 &      5.032(-06) & 1.231(-13) & 2.274(-05) & 2.548(-05) & 0.023 & 0.025\\
           &  7/2&    1897.70& E2 &      2.295(-06) & 5.639(-14) & 1.305(-05) & 1.461(-05) & 0.022 & 0.016\\
            &     &           & M1 &      7.864(-11) & 1.676(-16) & 3.880(-08) & 4.151(-08) & \\
           &  5/2&    1892.58& E2 &      6.260(-07) & 1.550(-14) & 4.812(-06) & 5.369(-06) & 0.022 & 0.002\\
           &     &           & M1 &      4.785(-08) & 1.022(-13) & 3.173(-05) & 3.388(-05) & \\
           &  3/2&    1892.61& E2 &      7.415(-08) & 1.836(-15) & 8.550(-07) & 9.553(-07) & 0.022 & 0.003\\
           &     &           & M1 &      3.149(-09) & 6.728(-15) & 3.132(-06) & 3.348(-06) & \\
${\!^2\!H}$&  9/2&    1783.86& E2 &      9.799(-09) & 2.898(-16) & 6.075(-08) & 7.040(-08) & 0.029 & 0.113\\
$\,_5\,{\!^2\!G}$&  7/2&    1741.92& E2 & 6.819(-10) & 2.166(-17) & 5.953(-09) & 6.962(-09) & 0.031 & 0.303\\
           &     &           & M1 &      1.319(-14) & 3.062(-20) & 8.413(-12) & 9.242(-12) & \\
           &  9/2&    1732.47& E2 &      3.157(-08) & 1.019(-15) & 2.265(-07) & 2.637(-07) & 0.030 & 0.136\\
$\,_5\,{\!^2\!F}$&  5/2&    1635.15& E2 &        2.997(-10) & 1.151(-17) & 4.786(-09) & 5.508(-09) & 0.028 & 0.127\\
           &     &           & M1 &      1.421(-12) & 3.514(-18) & 1.461(-09) & 1.590(-09) & \\
           &  7/2&    1632.54& E2 &      6.856(-09) & 2.646(-16) & 8.277(-08) & 9.525(-08) & 0.028 & 0.285\\
           &     &           & M1 &      5.094(-12) & 1.262(-17) & 3.947(-09) & 4.294(-09) & \\
$_3\,{\!^2\!D}$&  3/2&    1349.59& E2 &  1.282(-08) & 8.756(-16) & 8.017(-07) & 8.892(-07) & 0.021 & 0.381\\
           &     &           & M1 &      4.080(-12) & 1.222(-17) & 1.119(-08) & 1.191(-08) & \\
           &  5/2&    1349.29& E2 &      4.692(-08) & 3.207(-15) & 1.958(-06) & 2.175(-06) & 0.021 & 0.379\\
           &     &           & M1 &      7.206(-11) & 2.160(-16) & 1.319(-07) & 1.405(-07) & \\
$\,_3\,{\!^2\!G}$&  9/2&    1206.35&E2&  2.668(-07) & 2.552(-14) & 1.170(-05) & 1.330(-05) & 0.025 & 0.111\\
           &  7/2    &  1026.31  & E2 &  2.147(-08) & 2.054(-15) & 1.177(-06) & 1.338(-06) & 0.025 & 0.108\\
           &  &    1206.35& M1 &         9.927(-16) & 3.328(-21) & 1.907(-12) & 2.059(-12) & \\
${\!^2\!P}$ &  3/2&     998.82& E2 &     3.434(-07) & 5.786(-14) & 9.672(-05) & 1.085(-04) & 0.023 & 0.149\\
           &     &           & M1 &      1.911(-10) & 7.738(-16) & 1.293(-06) & 1.386(-06) & \\
           &  1/2&     998.74& E2 &      9.615(-08) & 1.621(-14) & 5.537(-05) & 6.209(-05) & 0.023 & 0.149\\
\hline
\end{tabular}
\end{minipage}
\end{table*}

What may be important for the accuracy of the wave function is the separation
of levels of the same $J$ and parity.  Table \ref{separation} shows the 
ten lowest observed $J=5/2$ levels and, for each level except the lowest,
the difference in energy of the level from the one immediately preceding it
in the table.  This difference represents the separation between levels of
the same $J$.
 Also reported is the ratio of the present and observed separation.
 The greatest deviation from
unity (with a ratio of 0.74) is for the separation between the 
\Term 4 G {5/2}/ and the \Term 4 P {5/2}/ levels:
the calculated separation of 2174 cm$^{-1}$ is too small compared with the
observed separation of 2953 cm$^{-1}$.
The second largest deviation from unity (with a ratio of 1.21)
is for the separation of the  \sterm 2 F 3/ and \Term 4 F {}/ $J=5/2$ levels. 
The important separations are those where configuration mixing occurs. 
A more detailed analysis of the wave function composition of
the ground state shows that the largest admixture is \Term 4 P {5/2}/, 
and that the \Term 6 S {5/2}/-\Term 4 P {5/2}/ energy separation
is accurate to 1.3~\%; the largest admixture
to \Term 4 G {5/2}/ is from \Term {2,4} F {5/2}/ and the energy separation
of the latter two with respect to \Term 4 G {5/2}/ has similar accuracy.
This suggests that the compositions of the \Term 6 S {5/2}/ and
\Term 4 G {5/2}/ wave functions are reliable.
Table \ref{spectrum} shows the larger admixture of \Term 4 D {5/2}/
in the \Term 4 P {5/2}/ wave function and vice versa.  The separation of
these energy levels is in error by 15~\%.  It is reasonable to assume
that the energy adjusted results from the MCHF calculation yield the more
accurate transition probability when this mixing is important.

\begin{table*}
 \centering
 \begin{minipage}{140mm}
\caption{Comparison of E2 and M1  transition rates $A_{J'J}$ (in
s$^{-1}$) for transitions to the ground state
for different theories: 1) from
\citet{garstang},  2) from Raassen \& Uylings (1997, private communication)
3) from \citet{nahar},  \citet{bp},  and
 present normalized (N) and unnormalized (U) MCDHF values.
}
\label{compare-E2}
\begin{tabular}{@{} r r r l l l l l l l l @{}}
  \hline
LS & $J$ & Obs $\Delta E$&
\multicolumn{3}{c}{Semi-empirical}&   {\small MCHF} &
\multicolumn{2}{c}{ Present Calculation}\\
\cline{4-6} \cline{8-9} \\
 & & & \multicolumn{1}{c}{1)} & \multicolumn{1}{c}{2)} &
  \multicolumn{1}{c}{3)} &\ &
(N) & (U)  \\
 \hline
\multicolumn{8}{l}{E2 transitions}\\
 \qG  & 9/2 & 32292.8 & $<$1.0(-09)& 6.92(-12) &3.06(-10) & 1.85(-09) & 9.29(-09) & 1.12(-08) \\
      & 5/2 &  32301.2 & $<$1.0(-09)& 1.95(-08)&7.64(-09) & 2.30(-08)& 1.33(-08) & 1.61(-08) \\
      & 7/2 &  32305.7 & $<$1.0(-09)& 3.18(-08)&1.19(-08) &3.32(-08) & 1.19(-08) & 1.44(-08)\\
 \qP  & 5/2 &  35253.8 & 3.9(-05)& 4.26(-05) & 6.28(-05) & 4.02(-05) & 3.93(-05) & 4.19(-05)\\
      & 3/2 &  35333.0 & 1.5(-05)& 1.99(-05) & 3.52(-05) & 1.21(-05) & 2.29(-05) & 2.44(-05) \\
      & 1/2 &  35406.6 & very small    & 1.67(-06) & 7.11(-06) & 1.47(-05) & 5.52(-06) & 5.87(-06) \\
 \qD  & 7/2 &  38779.4 & 1.1(-03)& 1.13(-03) & 1.22(-03) & 1.27(-03) & 1.10(-03) & 1.26(-03) \\
      & 1/2 &  38897.7 & 1.8(-04)& 1.97(-04) & 2.00(-04) & 2.17(-04) & 1.83(-04) & 2.08(-04) \\
      & 5/2 &  38935.1 & 1.0(-03)& 1.09(-03) & 1.15(-03) & 1.23(-03) & 1.08(-03) & 1.22(-03) \\
      & 3/2 &  38938.2 & 6.2(-04)& 6.72(-04) & 6.80(-04) & 7.33(-04) & 6.31(-04) & 7.12(-04) \\
\multicolumn{8}{l}{M1 transitions}\\
 \qG  & 5/2 &  32301.2 & 1.0(-05)& 1.53(-05) & 2.16(-06) & 2.16(-05) & 2.12(-05) & 2.38(-05) \\
      & 7/2 & 32305.7 & $<$1.0(-07)& 3.78(-08)& 8.29(-09) &5.72(-08)& 2.14(-08) & 2.40(-08)\\
 \qP  & 5/2 & 35253.8 & 1.4&       1.42       & 1.21      & 1.56      & 1.31 & 1.36 \\
      & 3/2 & 35333.3 & 8.8(-01)& 9.23(-01) & 7.92(-01)   & 1.02 & 8.49(-01) & 8.83(-01) \\
 \qD  & 7/2 & 38779.4 & 2.0(-04)& 5.90(-04) & 3.33(-04)   & 7.67(-04) & 3.08(-04) & 3.34(-04)\\
      & 5/2 & 38935.1 & 5.1(-02)& 5.69(-02) & 1.11(-02) & 6.64(-02) & 3.29(-02) & 3.56(-02) \\
      & 3/2 & 38938.2 & 3.8(-02)& 2.75(-02) & 3.56(-02) & 3.33(-02) & 1.58(-02) & 1.71(-02) \\
\hline
\end{tabular}
\end{minipage}
\end{table*}

Table \ref{excitation} reports transition probability data for E2 and M1 transitions to the 
ground state for all levels considered in this paper except for a few 
weak lines from $J=1/2$ levels.  Included in this table is the observed
wavelength (in vacuum) and the computed line strength $S$, 
the weighted oscillator
strength $gf$, the transition rate $A_{ki}(N)$ normalized to the 
observed wavelength and the {\it ab initio} unnormalized $A_{ki}(U)$ value,
both in $s^{-1}$. Thus the normalized value is obtained from the computed
line strength and the observed wavelength.
All these values are computed in the length form.
 Also included are ``indicators of accuracy''~\citep{asos9}.  The factor
$\delta E = \Delta E(calc)/\Delta E(obs) -1$ is a measure of the accuracy
of the transition energy and is always positive for the transitions 
considered in this table.  Consequently, normalization reduces all these
 transition rates.
The factor $$\delta T = |S(\mbox{length})-S(\mbox{velocity})|/ 
\max(S(\mbox{length}), S(\mbox{velocity}))$$ 
is a measure of the agreement in length and velocity 
values of the line strength $S$ for E2 transitions.  Because all of the 
transitions in this
table are $LS$ forbidden with contributions arising from small components 
of the wave function, some $\delta T$ factors are relatively large. 
Generally the length form of the line strength is the more stable value
as more correlation is included in the calculation.   

A complete list of similar data for all multiplets
between different $LS$ terms of $3d^5$ is provided in Appendix A (for E2) and
Appendix B (for M1) transitions.
Not included are transitions between levels within an $LS$ term for 
which the transition energy is less than 100 cm$^{-1}$.

\section{Comparison of transition probabilities}

\begin{table*}
 \centering
 \begin{minipage}{140mm}
\caption{Comparison of MCHF Breit-Pauli and present normalized transition rates
$A_{J'J}$ (in s$^{-1}$)
for $LS$ allowed  E2 and M1  transitions between excited states.
}
\label{compare-excited}
\begin{tabular}{@{} r r r r r l l l l r r @{}}
  \hline
\multicolumn{4}{l}{Transition} &  & 
\multicolumn{2}{l}{E2 transition} & & \multicolumn{2}{l}{M1 transition}\\
\cline{6-7}\cline{9-10}
 $LS$ & $L'S'$ & $J$ & $J'$  & & MCHF & Present & & MCHF & Present\\
 \hline
\hline
\dI & \sterm 2 G 3/ & 11/2 & 7/2 && 5.00      & 4.92 && \\
    &               &      & 9/2 && 1.50(-01) & 1.47(-01) && 2.24(-05) & 2.28(-05) \\
    &               & 13/2 & 9/2 && 4.82      & 4.78 \\
    & \dH           & 11/2 & 9/2 && 5.77(-04) & 5.67(-04) && 8.27(-02) & 8.24(-02) \\
    &               &      &11/2 && 2.58(-05) & 2.43(-05) && 2.05(-01) & 1.92(-01) \\
    &               & 13/2 & 9/2 && 1.52(-05) & 1.71(-05) && \\
    &               &      &11/2 && 6.60(-04) & 6.54(-04) && 1.07(-01) & 1.03(-01) \\
\dF & \dP           &  7/2 & 3/2 && 7.33      & 7.14 \\
    &               &  5/2 & 1/2 && 8.86     & 8.63 \\
    &               &      & 3/2 && 1.22      & 1.18      && 9.04(-05) & 1.10(-04)\\
\qG & \qF  & 5/2 & 3/2 && 2.00(-01) & 1.93(-01) && 1.59(-01) & 1.38(-01) \\
    &      &     & 5/2 && 6.36(-02) & 6.39(-02) && 2.91(-01) & 2.05(-01) \\
    &      &     & 7/2 && 6.69(-03) & 6.52(-03) && 1.58(-02) & 1.27(-02) \\
    &      &     & 9/2 && 8.95(-05) & 8.96(-05) \\
    &      & 7/2 & 3/2 && 8.00(-02) & 7.85(-02) \\
    &      &     & 5/2 && 1.10(-01) & 1.21(-01) && 2.11(-02) & 2.67(-03) \\
    &      &     & 7/2 && 7.17(-02) & 7.03(-02) && 1.56(-01) & 1.29(-01) \\
    &      &     & 9/2 && 3.91(-03) & 3.89(-03) && 1.25(-02) & 9.85(-03) \\
    &      & 9/2 & 5/2 && 6.00(-02) & 6.57(-02) \\
    &      &     & 7/2 && 1.55(-01) & 1.53(-01) && 5.12(-02) & 3.75(-02) \\
    &      &     & 9/2 && 4.71(-02) & 4.66(-02) && 7.31(-02) & 6.40(-02)\\
    &      & 11/2& 7/2 && 4.33(-02)& 4.26(-02) \\
    &      &     & 9/2 && 2.23(-01) & 2.20(-01) && 1.60(-01) & 1.36(-01)\\
\hline
\end{tabular}
\end{minipage}
\end{table*}

In order to further assess the accuracy of these results, 
we compare in Table \ref{compare-E2} the present transition probabilities between
\Term 6 S {}/ and the \Term 4 G {}/,  \Term 4 P {}/, and \Term 4 D {}/ terms
with previously published values.
Included are Garstang's early calculations, results
from  the semi-empirical orthogonal operator method
\citep[][1997 -- private communication]{raassen}, 
the one-body Breit-Pauli results reported by \citep{nahar},
the two-body  MCHF~\citep{bp} Breit-Pauli values, 
as well as the present normalized (N) and unnormalized (U) values.
With only one exception, namely 
the transition from \Term 4 G {9/2}/,  the process of normalizing to
the observed wavelength has reduced the transition rate to values below
those reported for MCHF previously. Since the 
separation of \Term 4 P {}/ and \Term 4 D {}/ is too large
(see Table \ref{separation}) the mixing of these terms is less than in MCHF.
The latter calculation used the $LS$ term energy corrections
to improve the wave function composition, a correction we were not able to
perform in the present work.
For transitions from \Term 4 G {}/ levels, the \citet{nahar} values are
considerably smaller, possibly because of the restricted Breit-Pauli operators
that were included.  As mentioned earlier, for these levels the Breit correction
was needed to produce levels in the observed order. The low-order version 
of this correction consists of the two-body Breit-Pauli operators 
that were omitted in 
Nahar's calculations.
What is striking is that, when there is excellent agreement among the last
three columns (and possibly others), there is also agreement within 10~\% 
with Garstang's 1958 result. Examples are some of the transitions from
\Term 4 P {}/.  Exceptions tend to be the very small values for which the 
transition rate is   $< 10^{-9}$ s$^{-1}$.

Table \ref{compare-excited} compares $LS$ allowed values for E2 and M1 transitions.
All results are now the normalized values. 
In the majority of the cases there is agreement to within 3~\% between the two
theories over a wide range of transition rates.  

Both the Breit-Pauli MCHF method with term energy corrections 
and the present fully relativistic MCDHF method have points in their favour.  In
order to establish a set of ``best'' values, the two sets were merged and
agreement in length and velocity used to select between them. Exceptions
are the 1/2 - 1/2 transitions for which E2 transitions are not allowed.  
In this case the value selected was from the same calculation as for
a 3/2 - 1/2 transition of the same multiplet.  These values are reported in
Appendix C.  Transitions are identified by
the index of the lower and upper level (as given in 
table \ref{spectrum}) and the data include the wavelength (in vacuum), 
$A_{ki}(\mbox{E2})$,
$A_{ki}(\mbox{M1})$, $A_{ki}(\mbox{Total})$ and a symbol for the source: {\tt B} for the
earlier Breit-Pauli MCHF data and {\tt G} for the present results 
obtained using the GRASP program. 
A total of 315 {\tt G} transitions were selected 
and 150 {\tt B} transitions.  It should be noted that the agreement in the 
length and velocity values is not a definitive indicator of accuracy, 
but is a reasonable one.  All the present normalized values of Table 4 were
selected as the ``best'' by this process.

The publications over the last decade reporting transition data
for transitions between levels of $3d^5$ in {[Fe~\sc{iv}]}
illustrate the difficulty of establishing the accuracy of transition rates in
complex cases.  Garstang computed many values fifty years ago.  Progress has
been made but agreement between the MCHF Breit-Pauli and 
MCDHF values for more transitions would be desirable.

In Appendix C
we have listed our recommended transition rates for both
E2 amd M1 transitions as well as their sum for the convenience of the
community. All appendices are in {\tt ASCII} form for ease in data processing.

\section*{Acknowledgments}

Support for this publication was provided by NASA through
Program number HST-AR-10973.01-A (PI RR) from  the
Space Telescope Science Institute, which is operated by the
Association of Universities for Research in Astronomy,
under NASA contract NAS5-26555.
MR acknowledges support from Mexican CONACYT project 50359-F.


\scriptsize
\appendix

\section{E2 transition data and accuracy}



\clearpage
\section{M1 transition data and accuracy}

%

\clearpage
\footnotesize

\section{Recommended values from MCDHF (G) and MCHF (B) }

%

\bsp

\label{lastpage}
\end{document}